\def\year{2021}\relax
\documentclass[letterpaper]{article} 
\usepackage{float} 

\usepackage{aaai21}  
\usepackage{times}  
\usepackage{helvet} 
\usepackage{courier}  
\usepackage[hyphens]{url}  
\usepackage{graphicx} 
\usepackage{subfig}
\usepackage{natbib}  
\usepackage{caption} 
\title{Quantifying Uncertainty for Machine Learning Based Diagnostic}
\author {
    Owen Convery, \textsuperscript{\rm 1}
    Lewis Smith, \textsuperscript{\rm 2}
    Yarin Gal, \textsuperscript{\rm 2}
    Adi Hanuka, \textsuperscript{\rm 1} \\
}
\affiliations {
    \textsuperscript{\rm 1} SLAC National Accelerator Laboratory, Menlo Park, CA 94025, USA \\
    \textsuperscript{\rm 2} Department of Computer Science, University of Oxford, UK \\
    adiha@slac.stanford.edu
}

\begin{document}

\maketitle

\begin{abstract}
Virtual Diagnostic (VD) is a deep learning tool that can be used to predict a diagnostic output.
VDs are especially useful in systems where measuring the output is invasive, limited, costly or runs the risk of damaging the output.
Given a prediction, it is necessary to relay how reliable that prediction is. This is known as `uncertainty quantification' of a prediction. IN this paper, we use ensemble methods and quantile regression neural networks to explore different ways of creating and analyzing prediction's uncertainty on experimental data from the Linac Coherent Light Source at SLAC. We aim to accurately and confidently predict the current profile or longitudinal phase space images of the electron beam.
The ability to make informed decisions under uncertainty is crucial for reliable  deployment of deep learning tools on safety-critical systems as particle accelerators.
\end{abstract}

\section{ Introduction }

Particle accelerators serve a wide variety of applications ranging from  chemistry, physics to biology experiments. Those experiments require increased accuracy of diagnostics tools to measure the electron beam properties during its acceleration, transport and delivery to users.
Current state-of-the-art diagnostics \cite{XTCAV} have limited applicability as the complexity of the experiments grows.
Virtual diagnostic (VD) tools provide a shot-to-shot non-invasive measurement of the beam in cases where the diagnostic has limited resolution or unavailable.
VDs have the potential to be useful also in experiments design, setup and optimization while saving valuable operation time. They could also aid in  interpreting experimental results, especially in cases in which current diagnostics cannot provide necessary information.
In this extended abstract, we apply deep learning tools to provide confidence interval of the virtual diagnostic prediction using experimental data from the Linac Coherent Light Source (LCLS) at SLAC.

Current VD provides predictive models based on training a neural network mapping between non-invasive diagnostic input to invasive output measurement \citep{Emma2018, hanuka2020accurate}. This type of mapping is known as supervised regression. Previous work has demonstrated VD to predict the electron beam current profile and Longitudinal Phase Space (LPS) distribution \cite{DESY:lps} along the accelerator using either scalar controls \cite{Emma2018} or spectral information \cite{hanuka2020accurate} as the non-invasive input to the VD.
However, an essential component in deploying the virtual diagnostic tool is to quantify the confidence in the prediction, i.e. estimate an interval presenting the uncertainty in prediction. This becomes critical in systems such as particle accelerators, where safety is of highest concern. In addition, users must have a reliable presentation of uncertainty given a set of inputs in order to make informed decisions.

\section{ Particle Accelerators and the Data Set }

High brightness beam linacs typically operate in single-pass, multi-stage configurations where a high-density electron beam formed in the RF gun is accelerated and manipulated prior to delivery to users in an experimental station. An example of such a facility is the LCLS XFEL at SLAC where the electron beam traverses through an undulator, and emits coherent X-ray pulses. Typically, longitudinal phase space (LPS) is destructively measured by X-band transverse deflecting cavity (XTCAV) \cite{XTCAV}.
In this abstract we used a longitudinal phase space (LPS) data set measured at the XTCAV, as the outputs, and the corresponding spectral information, as can be collected by the IR spectrometer, as an input.

\begin{figure}[h!]
    \centering
    \subfloat{{\includegraphics[width=3.75cm]{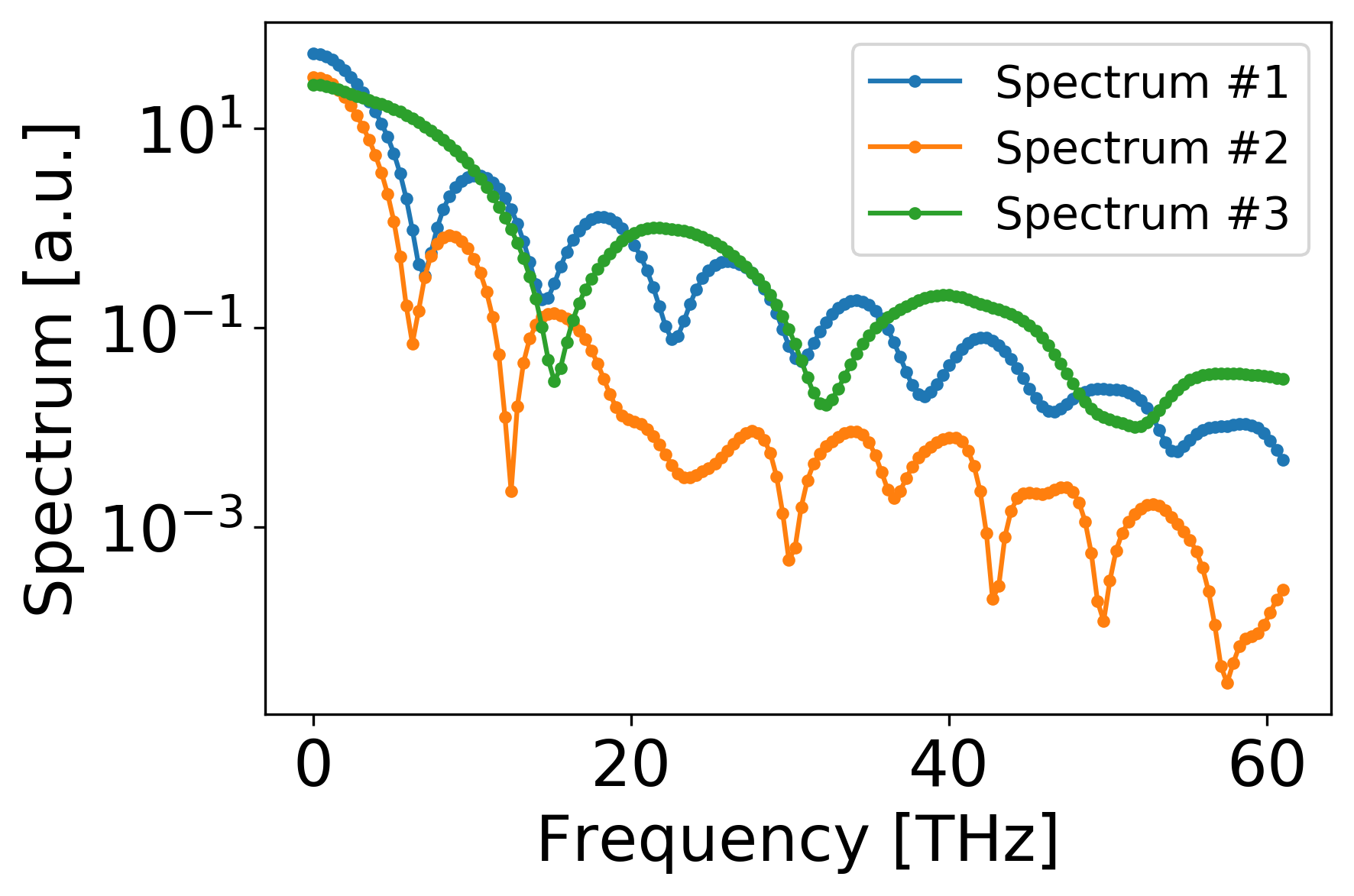} }}%
    \qquad
    \subfloat{{\includegraphics[width=3.75cm]{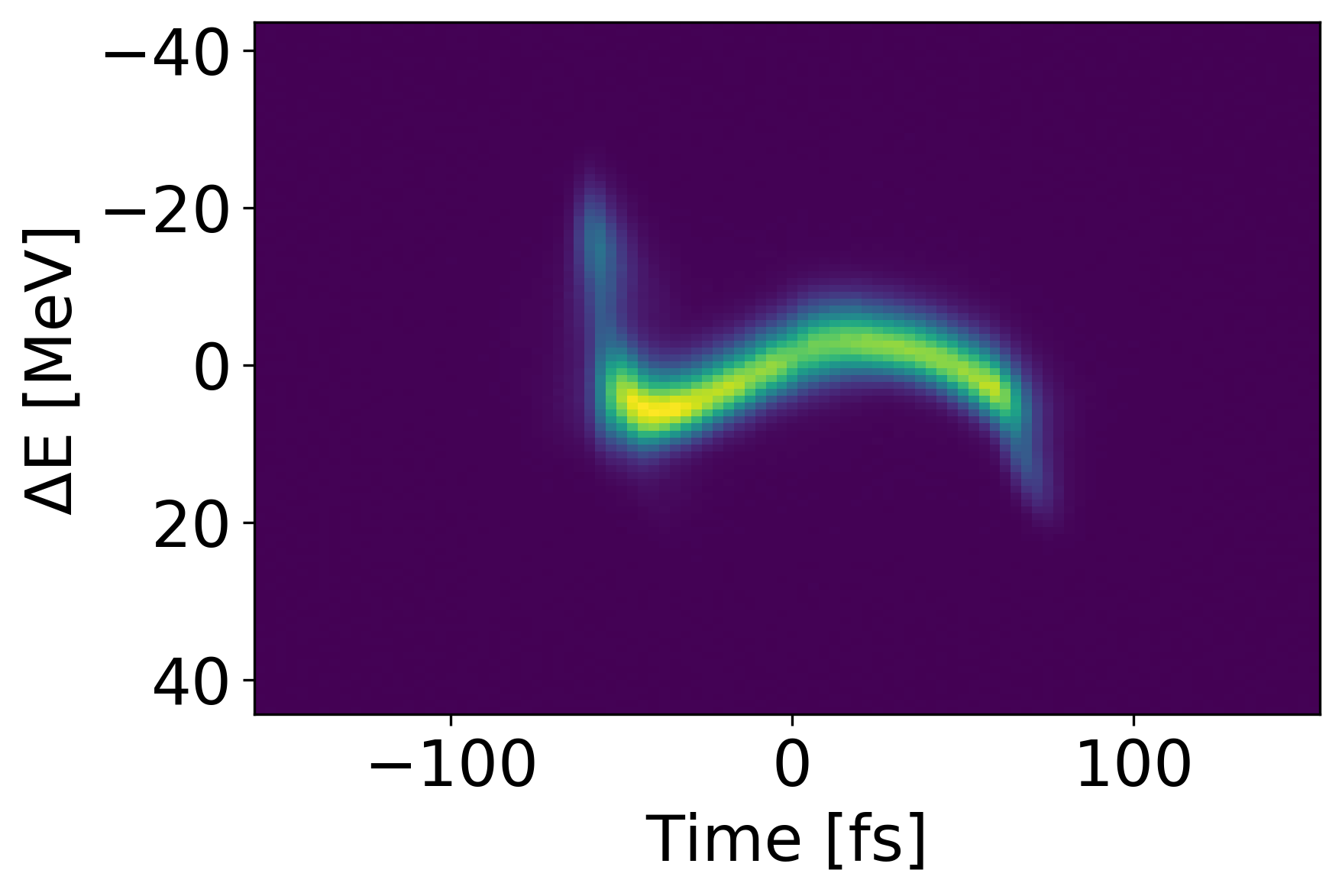}}}%
    \qquad
    \subfloat{{\includegraphics[width=3.75cm]{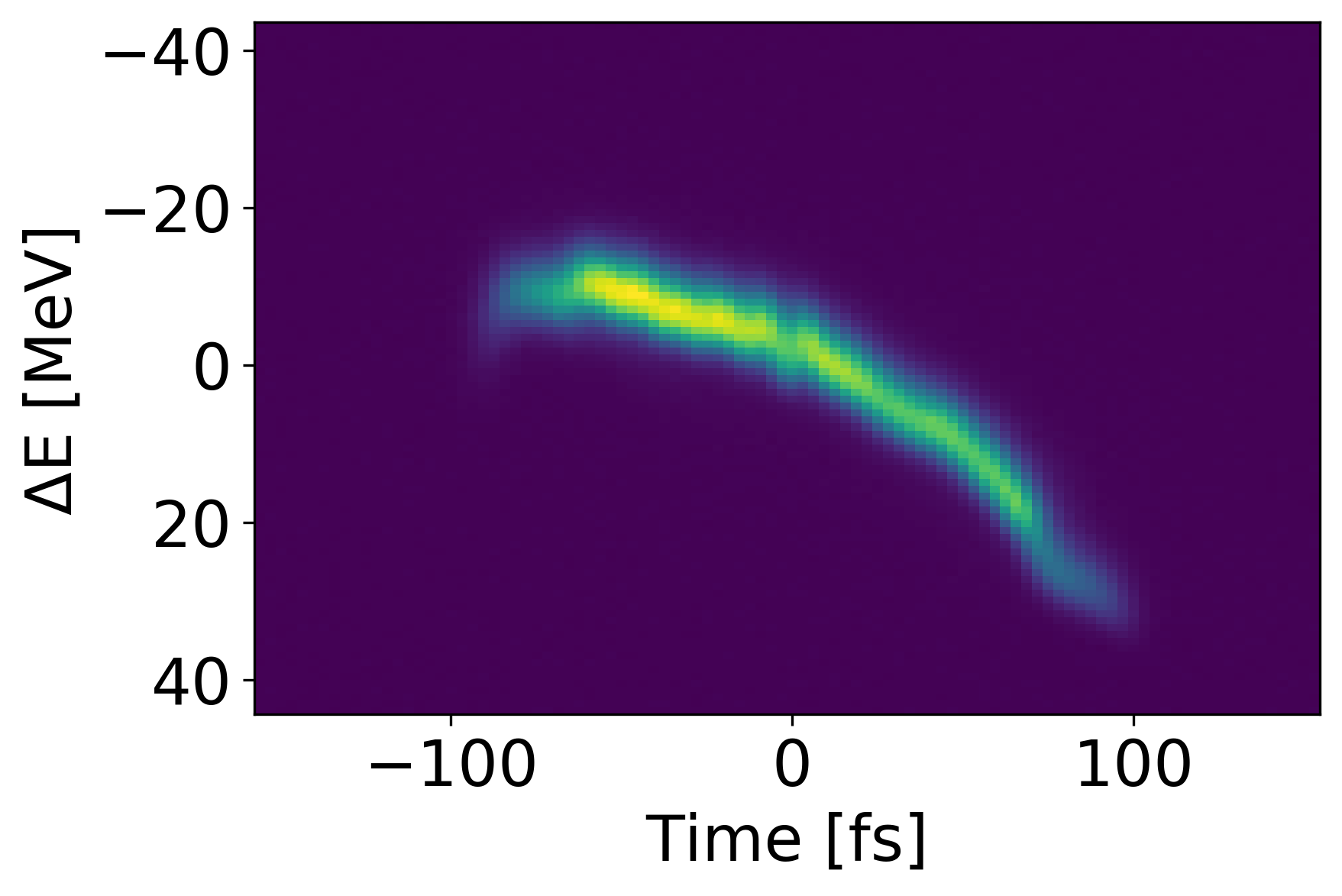} }}%
    \qquad
    \subfloat{{\includegraphics[width=3.75cm]{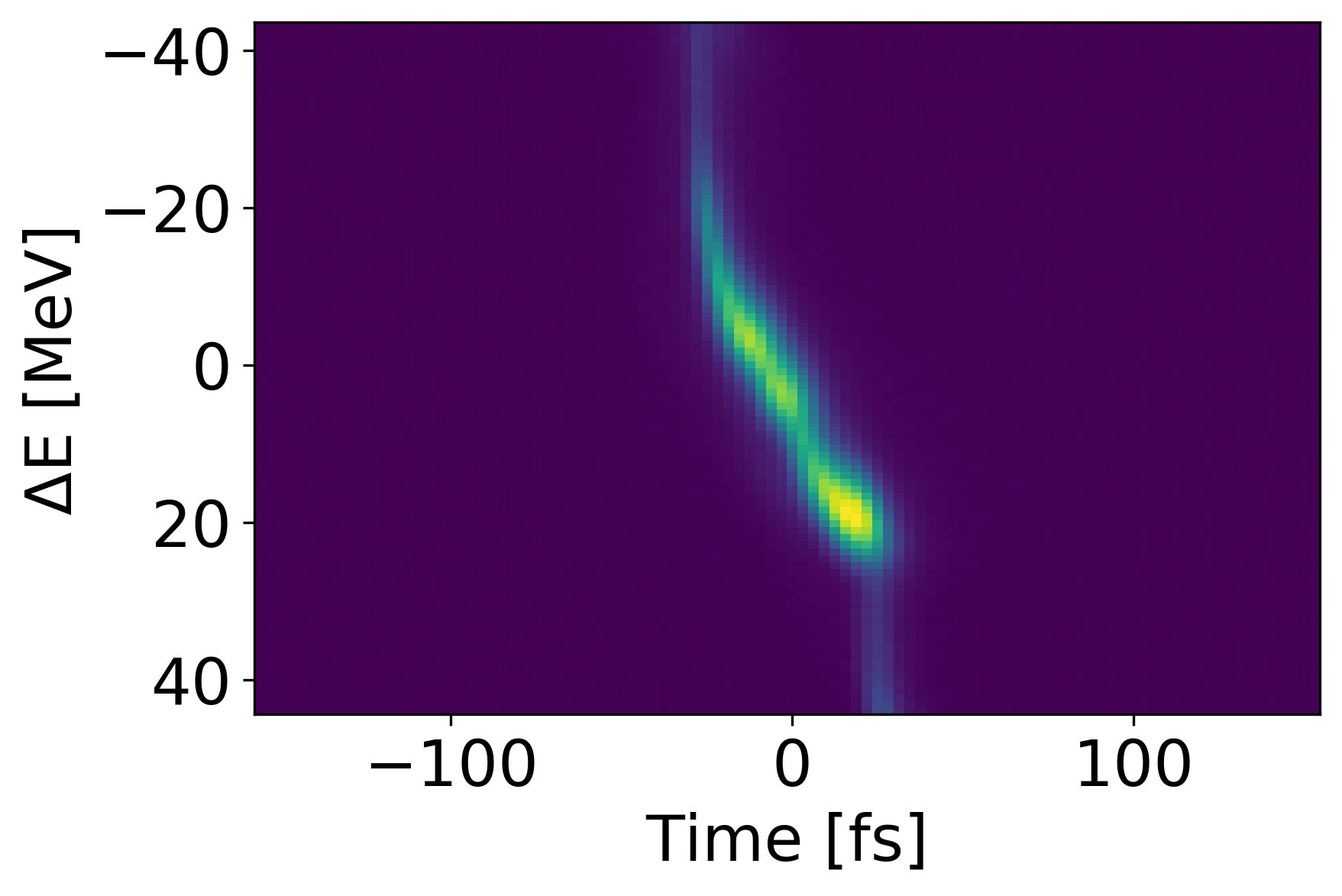} }}%
    \caption{Example of Spectral VD input and longitudinal phase space outputs}%
    \label{fig:VD_input_output}%
\end{figure}

\section{ Methods and Metrics}

Deep ensembles are ensembles of neural networks, each initialised differently and optimised to convergence independently of the others. 
Ref. \cite{lakshminarayanan2017simple} has shown empirically that explicit ensembling {can} result in improved uncertainty estimates when they used large neural networks with non-convex loss surfaces.

The predicted current profile for a test shot $\vec{I}_{{\rm predicted}}=M^{-1}\sum_{m=1}^M \vec{i}_{{\rm predicted},m} $ is the mean prediction of $M$ neural network predictions $\vec{i}_{{\rm predicted},m} $.
The uncertainty is manifested as the standard deviation of the neural network predictions $\vec{\sigma}=\sqrt{M^{-1}\sum_{m=1}^M (\vec{i}_{{\rm predicted},m}-\vec{I}_{\rm predicted})^2}$.

The neural network (NN) architecture we used is a fully connected feed-forward NN composed of three hidden layers (200, 100, 50) with rectified linear unit activation function.
For training we used batch size of 32, 500 epochs and Adam optimizer with fixed learning rate of 0.001 in our experiments \cite{hanuka2020accurate}. 
Training the NNs with a Gaussian likelihood, i.e. to minimize the standard Mean Squared Error (MSE) loss function on a training set, yields symmetric uncertainty intervals. 
For all the examples presented we use the open source Keras and TensorFlow libraries to build and train the NN module \cite{keras,TF}. 

To evaluate the mean prediction of our VD, we use the standard mean squared error metrics (${\rm MSE} = {T^{-1}\sum_{t=1}^{T} {(I_{{\rm measured},t}-I_{{\rm predicted},t})}^{2}}$). 
To evaluate the uncertainty intervals provided by the predictive standard deviation, we use a custom accuracy metric:

\begin{equation}
    {{\rm Accuracy }=  \frac{\sum_{t=1}^{T}\alpha_t\cdot  I_{{\rm measured},t}^{2}}{ \sum_{t=1}^{T} I^{2}_{{\rm measured},t}}}
        \label{eq:accuracy}
\end{equation}

where ${\rm \alpha}_t =1$ if $I_{\rm lower}{,t} < I_{{\rm measured}{,t}} < I_{{\rm upper}{,t}}$ and 0 otherwise. We used bounds of $I_{{\rm predicted}{,t}}\pm2\sigma_{t}$ where $\sigma_t$ is predictive standard deviation at time $t$.

\section{ Results and Discussion }

\begin{figure}[h!]
    \centering
    \includegraphics[width = 8cm]{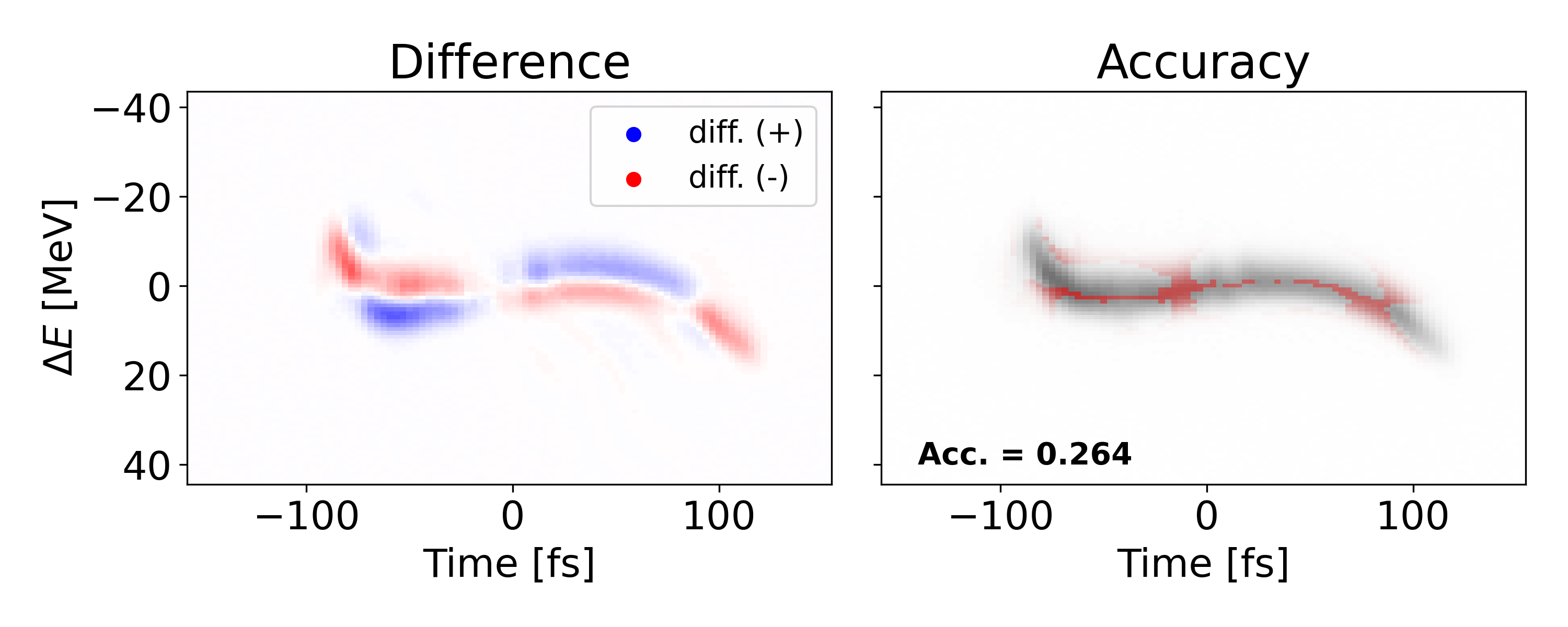}
    \caption{ A test shot with prevalent shape error }
    \label{fig:shape}
\end{figure}

The average MSE of the VD on the test set is 6.714e-04 with an accuracy of 0.538 using $\pm 2 \sigma$. 
Figure \ref{fig:shape} shows an example of a poor test shot with a few tools to analyze the VD performance. 
We present a comparison between the VD prediction and ground truth on the left, and the ground truth with red indicating where the prediction is accurate.

In the shown test shot we observe a large MSE of 1.096e-3 and a low accuracy of 0.208. 
The large MSE becomes obvious when we look at the second panel depicting the error.
In the case of this shot, the shape of the prediction is completely off. 
Looking at the right most panel, we see the prediction is the most uncertain near the tails of the signal and confident near the center.

\begin{figure}[!h]
    \centering
    \subfloat{{\includegraphics[width=8cm]{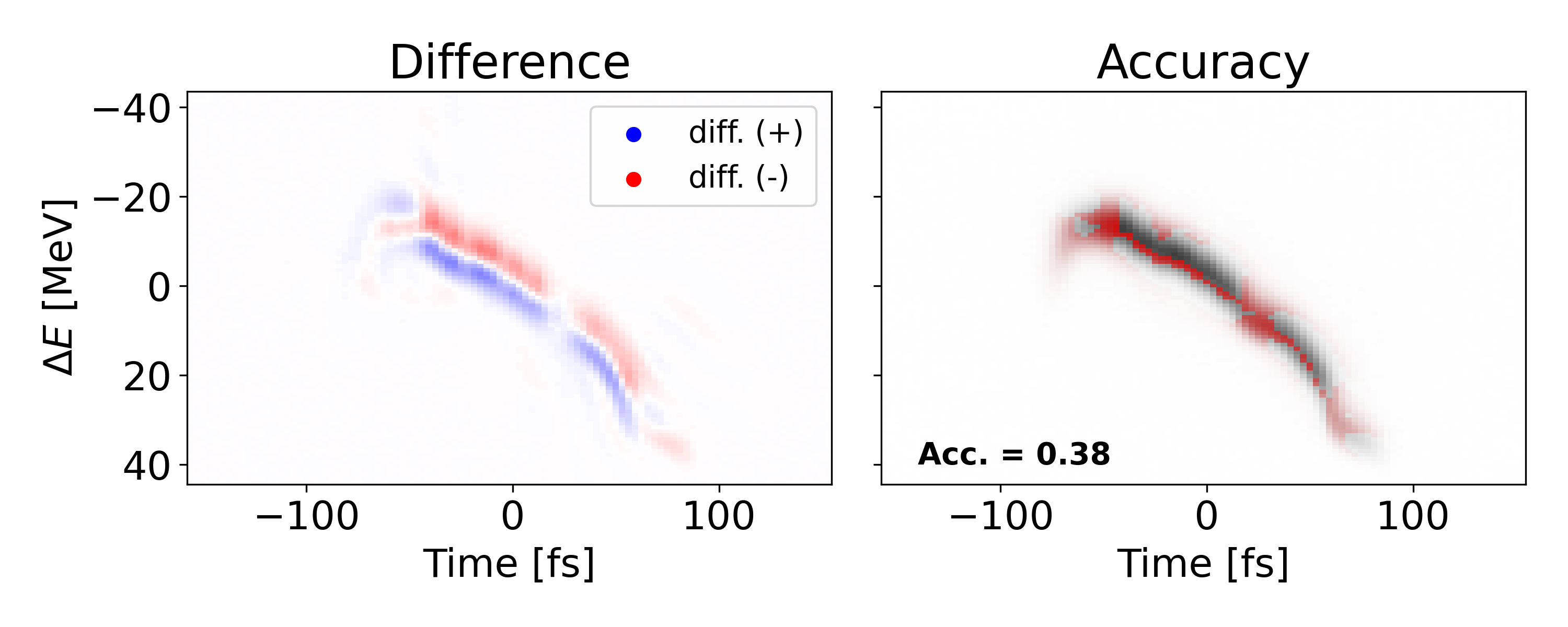} }}%
    \caption{ Difference and Accuracy before translation }
    \subfloat{{\includegraphics[width=8cm]{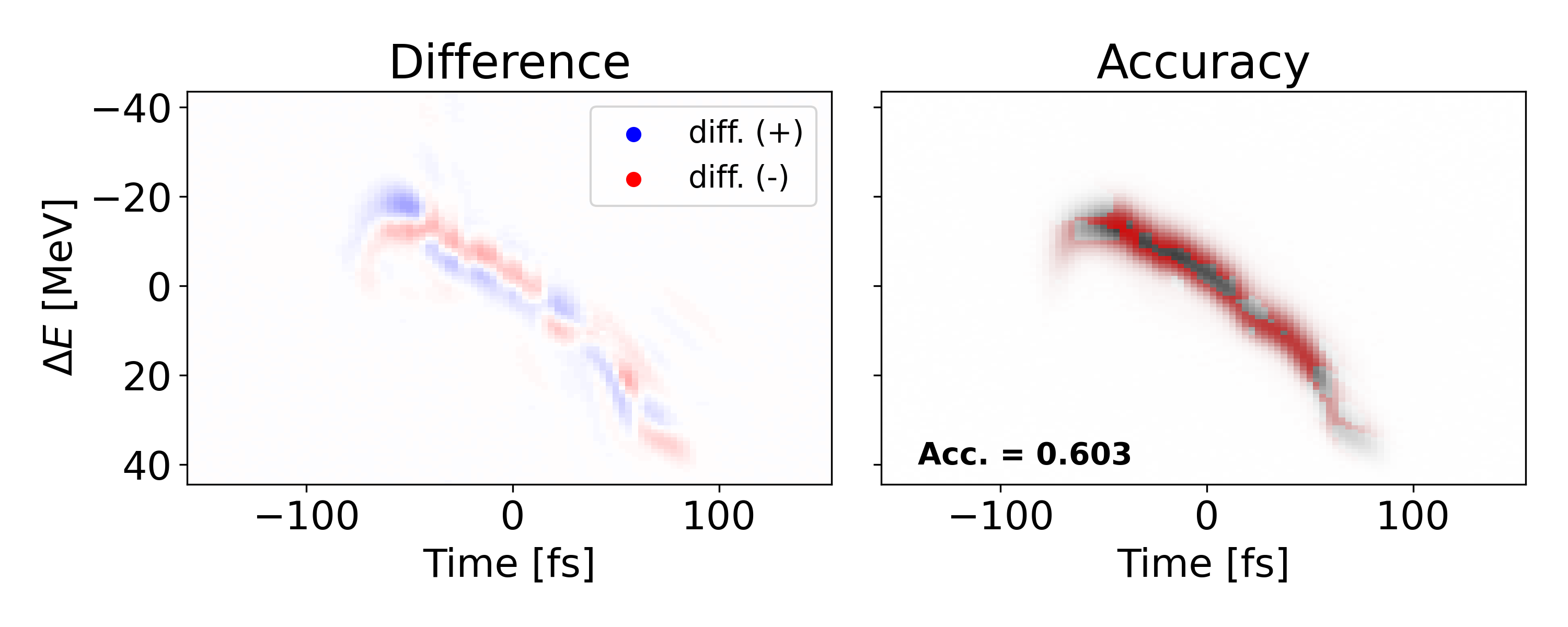} }}
    \caption{ Difference and Accuracy after translation }
\end{figure}

Another poor shot is shown above with an MSE of 2.737e-4 and accuracy of 0.383.
However, this lower performance is due to translational error, not shape error as seen in the previous example.
Since we mostly care about the shape of the signal, we can translate the prediction to match the "center of mass" for the measured value. 
Doing so yields an MSE of 8.396e-5 and and accuracy of 0.661 which indicates an improvement of 85.0$\%$ and 41.8$\%$ respectively.
Both types of error could potentially be reduced if spatial connectivity was leveraged in a more sophisticated network architecture.

\section{ Conclusions and Outlook }

In this abstract we have presented methods, metrics, and tools to present uncertainty quantification for standalone LPS images. 
Although looking at individual shots allows us pinpoint data set features and analyze problems with our VD, it does not give much insight into how the VD performs on the data set as a whole.
In future research, we will present tools to visualize uncertainty over an entire test set and allow users to make crucial decisions regarding rejection thresholds and machine operations.

\section{ Acknowledgments }

This work was supported by the Department of Energy, Laboratory Directed Research and Development program at SLAC National Accelerator Laboratory, under contract DE-AC02-76SF00515.

\bibliography{ref}

\end{document}